\documentclass[reprint,prl,twocolumn,showpacs,epsfig,longeq,superscriptaddress]{revtex4}
\usepackage{graphicx}
\usepackage{dcolumn}
\usepackage{amssymb}
\usepackage{bm}
\usepackage{epsfig}
\usepackage{psfrag}

\def\3dots{\:\raisebox{-0.5ex}{$\stackrel{\textstyle.}{:}$}\:}

\def\beq{\begin{equation}}
\def\eeq{\end{equation}}
\def\bea{\begin{eqnarray}}
\def\eea{\end{eqnarray}}
\def\mc{meridional circulation}
\def\pf{poloidal field}

\begin{document}

\title{The origin of grand minima in the sunspot cycle}

\author{Arnab Rai Choudhuri}
\email[Corresponding author email: ]{arnab@physics.iisc.ernet.in}
\affiliation{Department of Physics, Indian Institute of Science, Bangalore 560012, India}
\affiliation{National Astronomical Observatory of Japan, Mitaka, Tokyo 181-8588, Japan}

\author{Bidya Binay Karak}
\affiliation{Department of Physics, Indian Institute of Science, Bangalore 560012, India}
\pacs{}

\draft

\begin{abstract}
One of the most striking aspects of the 11-year sunspot cycle is that there have been times in the past when some cycles went missing, a most well-known example of this being the Maunder minimum during 1645---1715. Analyses of cosmogenic isotopes ($^{14}$C and $^{10}$Be) indicated that there were about 27 grand minima in the last 11,000~yr, implying that about 2.7\% of the solar cycles had conditions appropriate for forcing the Sun into grand minima. We address the question how grand minima are produced and specifically calculate the frequency of occurrence of grand minima from a theoretical dynamo model. We assume that fluctuations in the poloidal field generation mechanism and in the meridional circulation produce irregularities of sunspot cycles. Taking these fluctuations to be Gaussian and estimating the values of important parameters from the data of last 28 solar cycles, we show from our flux transport dynamo model that about 1--4\% of the sunspot cycles may have conditions suitable for inducing grand minima.
\end{abstract}

\maketitle

A few years after the initiation of telescopic observations of sunspots
in 1610, there was a period from 1645 to 1715 when very few sunspots
appeared on the face of the Sun. This period is known as the Maunder minimum \cite{eddy}.
Although reliable sunspot data did not exist before 1610, the solar activity at earlier times
can be studied from the analyses of the abundances of cosmogenic isotopes
like $^{14}$C in old tree rings \cite{solanki04,nagaya} and $^{10}$Be in polar ice \cite{beer90}.
When sunspots are absent, the magnetic field in the solar wind becomes weak and more
galactic cosmic rays can reach the Earth, producing more of such radioactive isotopes
in the atmosphere. Analyses of these isotopes indicate that there have been about
27 grand minima of different durations in the last 11,000 years \cite{usos07}.
Since there were about 1,000 solar cycles during this period, the occurrence of 27 grand minima implies
that about 2.7\% cycles had conditions appropriate for forcing
the Sun into grand minima. Furthermore this study showed that
the Sun was in the grand minima state for about 17\% of the time.
We also mention that there are evidences that some solar-like stars show grand
minima \cite{baliu}. Therefore it is very important to understand the physics of
the origin of the grand minima and the probability of occurrence of such grand minima.
Moreover, sunspot cycles have an important effect on the space environment and the Earth
climate system \cite{lean95,crowley00}. Therefore, understanding the grand minima is
also important to the space weather and Earth climate communities.

It appears that the most promising theoretical model for the solar cycle
at the present time is the flux transport dynamo model \cite{csd95, dc99, nc02}, which has
been reviewed recently by Charbonneau \cite{c10} and Choudhuri \cite{c11}. Let us
consider the question how irregularities arise in solar cycles. One
important assumption of the flux transport dynamo model is that the
poloidal field is produced by the Babcock--Leighton mechanism, in which
tilted bipolar sunspots give rise to the poloidal field after their decay.
The amount of poloidal field generated depends on the tilt angle of the
bipolar sunspots.  This tilt is produced by the action of the Coriolis
force acting on the magnetic flux tube rising through the solar convection
zone due to magnetic buoyancy \cite{dc93} and the average tilt at a solar latitude
is given by Joy's law.  However, there is a large scatter in the tilt angles
around the average given by Joy's law \cite{dasi10}, presumably due to the effects of
turbulence on the rising flux tubes \cite{lc02}.  Hence the Babcock--Leighton mechanism has an
inherent randomness because of which the poloidal field generated at the
end of a cycle would vary from one cycle to the other \cite{ccj07}. The second source
of irregularities in solar cycles comes from the fluctuations in the meridional
circulation of the Sun, which plays a crucial role in the flux transport
dynamo.  The period of the cycle in the theoretical model 
is approximately inversely proportional
to the amplitude of the meridional circulation \cite{dc99,ynm08}. Presumably the variations
in the periods of past cycles were produced primarily by variations in the 
meridional circulation. Although we have direct measurements of the
meridional circulation flow speed only during the last few years \cite{hr10,ulrich10}, the periods
of past cycles can be used to draw inferences about meridional circulation
variations in the past \cite{k10,kc11,pl08,lp09}. Such variations in meridional circulation also do cause variations
in the strengths of different cycles in addition to variations in their periods.  Suppose
the meridional circulation has become weaker than usual.  Then the period of
the cycle will be longer and diffusion will have more time to act on the
magnetic field, making the cycle weaker. If 
diffusion is assumed in the range $10^{12}$--$10^{13}$ cm$^2$ s$^{-1}$
consistent with mixing length arguments \cite{pa79}
 (which we do in our model), then
this effect overcomes the opposing effect of differential rotation also getting
more time to generate more toroidal field and the cycles become weaker when the
meridional circulation slows down \cite{ynm08,k10,kc11}. However, in the model 
developed by the HAO group, the diffusion is taken to be about 50 times smaller
than what we take \cite{dc99,dg06}, leading to the opposite effect of cycles 
getting stronger with slower meridional circulation due to the generation of
more toroidal field by differential rotation over a longer time. Several
arguments in favor of the higher diffusivity used by us are summarized in Sect.~5
of Jiang et al.\ \cite{jcc07}.

In order to model grand minima theoretically, we have to run our theoretical
flux transport dynamo model with fluctuations in poloidal field generation
and fluctuations in meridional circulation. Some studies have shown that
large fluctuations in poloidal field generation mechanism or large fluctuations in 
the meridional circulation can force 
the dynamo into intermittencies resembling grand minima 
\cite{c92,char04,ck09,usoskin09,pl11,pcb12}.
Karak \cite{k10} found that the dynamo is
driven into a grand minimum if the poloidal field at the end of a cycle and the 
meridional circulation at that time fall to sufficiently low values. Defining
a grand minimum as absence of sunspots for at least 20 years (the same definition as 
used by Usoskin et al.\ \cite{usos07} in estimating the number of grand minima in the
past 11,000 years), the shaded region of Fig.~1 indicates the combined
values of poloidal field and meridional circulation at the end of a cycle
necessary for forcing the dynamo into grand minima according to our theoretical
model.  The important question now is to estimate the probability of this
happening at the end of a cycle.  We estimate this probability in the
following way.
\begin{figure}[h]
\centerline{\includegraphics[width=0.55\textwidth]{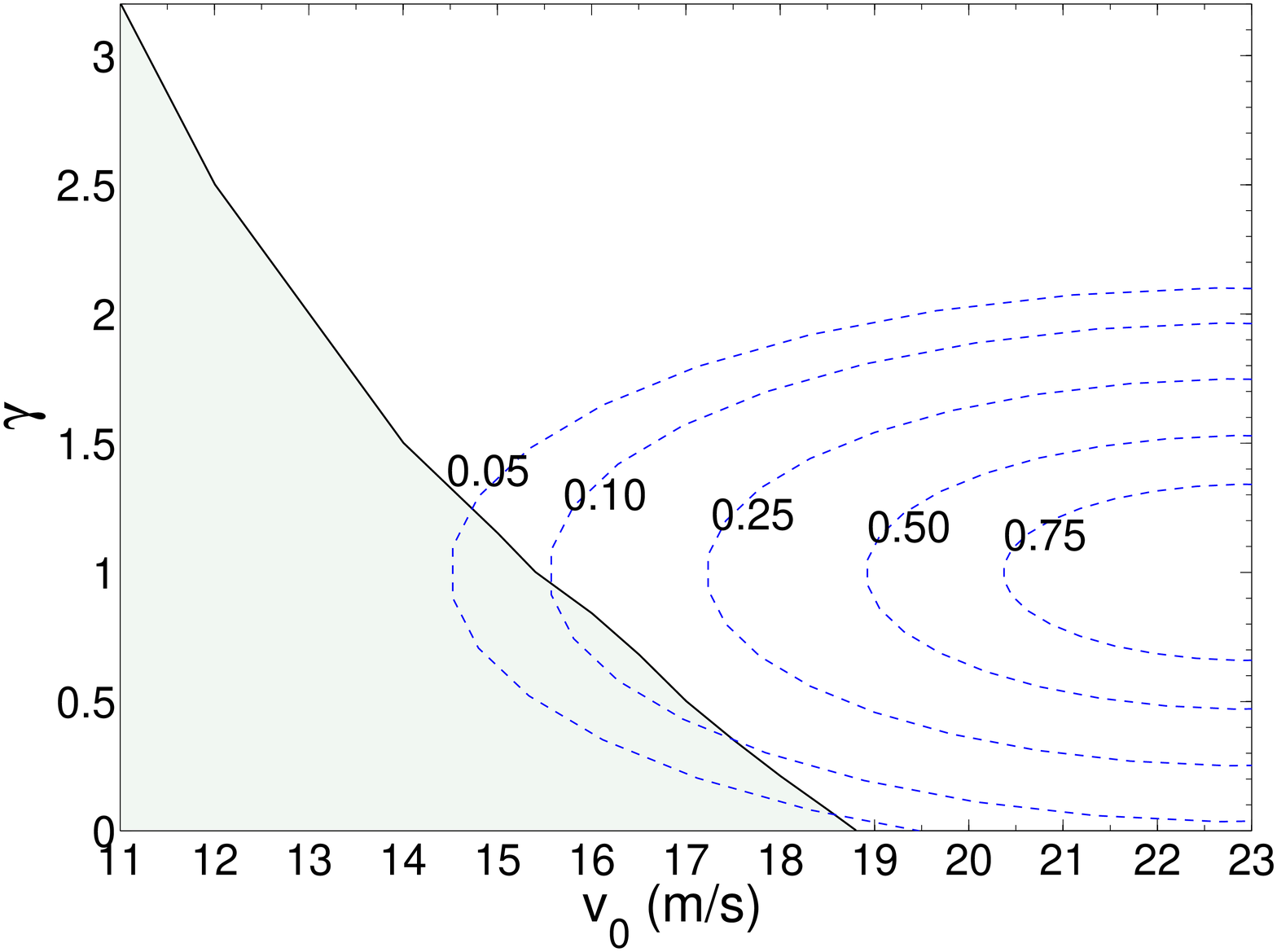}}
\caption{The solid line shows the values of the \mc\ amplitude $v_0$ and the poloidal
field scale factor $\gamma$ which produce grand minima of duration
$\sim 20$~yrs. The parameters lying in the shaded region produce grand minima of
longer duration. The dashed curves are the contours of the joint probability
$P(\gamma, v_0)$, with the values of $P(\gamma, v_0)$ (excluding the constant pre-factor) given in the plot.
The scale factor $\gamma$ is defined as the amplitude of poloidal field at the end of a cycle
divided by its value in the absence of fluctuations.  This figure is produced by using
our theoretical solar dynamo model (details given in Karak \cite{k10}).}
\label{parameters}
\end{figure}
\begin{figure}[h]
\centerline{\includegraphics[width=0.55\textwidth]{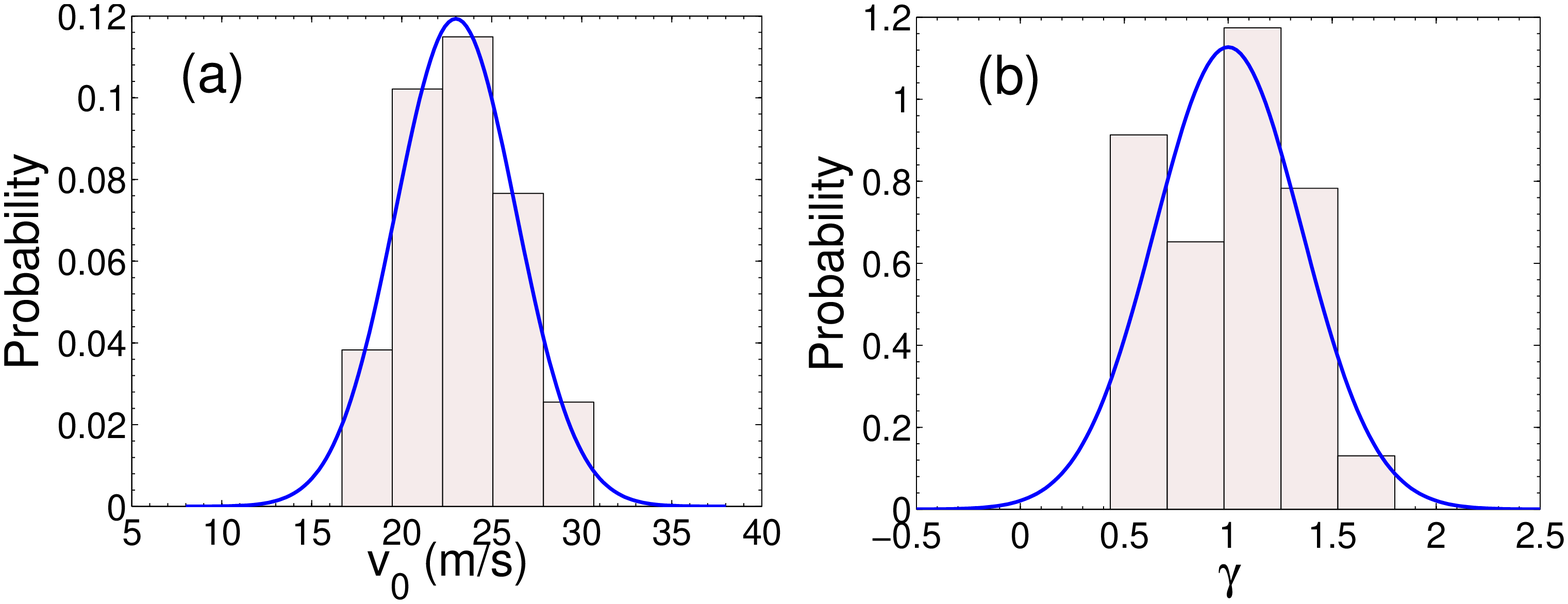}}
\caption{(a) Histogram of the \mc\ amplitude derived from the observed periods of last 28 solar cycles.
The solid curve is the Gaussian function with mean = 23 m~s$^{-1}$ and standard
deviation $\sigma_v$ = 3.34 m~s$^{-1}$. (b) Histogram of poloidal field scale factor
derived from the peak sunspot number of last 28 solar cycles. The solid curve is the
Gaussian function with mean = 1 and standard deviation $\sigma_{\gamma}$ = 0.35. The probabilities
plotted along the vertical axis in both (a) and (b) are obtained by dividing the number of data points in a data
bin by the total number of data points and the horizontal width of the data bin.}
\label{hist}
\end{figure}

Assuming that the inverse of the cycle period gives an approximate value of
the meridional circulation amplitude during that cycle (as suggested by
theoretical flux transport dynamo models \cite{dc99,ynm08}), Karak \& Choudhuri \cite{kc11}
concluded that the meridional circulation has changed randomly in the past
with a correlation time around 30--40~yr
(also see the similar study based on the low order dynamo model \cite{pl08,lp09}). 
Figure~2a shows a histogram
of the estimated values of the \mc\ amplitude during the last 28 cycles.  The
solid curve in Fig.~2a shows the Gaussian having the mean $\overline{v_0} =
23$ m s$^{-1}$ and the standard deviation $\sigma_v = 3.34$ m s$^{-1}$
calculated from the data presented in the histogram. We find that the
Gaussian is a reasonable fit, although we do not have any data points lying
way out in the Gaussian tail.  Jiang et al.\  \cite{jcc07} (also see \cite{dasi10}) pointed out on the basis
of observational data that there is a good correlation between the \pf\ at
the end of a cycle and the strength of the next cycle.  
Assuming a perfect correlation, we can use the strengths of following cycles
to obtain values $\gamma$ of \pf\ at the ends of previous cycles (scaled by taking their average
value as the unit). Figure~2b is a histogram
of the values $\gamma$ of \pf\ obtained in this way. It should, however,
be kept in mind that variations in the \mc\ also contribute to fluctuations
of cycle strengths \cite{baumann04,ynm08,k10,kc11} and fluctuations of \pf\ at the end of cycles obtained by
our method (as shown in Fig.~2b) is probably an over-estimate. The solid curve
in Fig.~2b is the Gaussian having the standard deviation $\sigma_{\gamma} 
= 0.35$ obtained from the data in the histogram 
(the mean is 1 by definition). We expect on general grounds
the distribution of $\gamma$ to follow a Gaussian, although we find that
the Gaussian fit in Fig.~2b is not as good a fit as in Fig.~2a. This is
not surprising given that our data set comprising of only 28 cycles 
is quite small.  If we assume that
the fluctuations in the amplitude $v_0$ of \mc\ at the solar surface and  
the fluctuations in $\gamma$ both follow Gaussian distributions, then we
can draw one obvious inference.  The joint probability that the \pf\ at the
end of a cycle lies in the range $\gamma$, $\gamma + d\gamma$ and the
amplitude of \mc\ at the same time lies in the range $v_0$, $v_0 +dv_0$ is given by
$$P (\gamma, v_0) d\gamma dv_0 = \frac{1}{\sigma_v \sqrt{2 \pi}} \exp\left[- \frac{(v_0 - \overline{v_0})^2}{2 \sigma_v^2}\right]$$ \\
$$\times \frac{1}{\sigma_{\gamma} \sqrt{2 \pi}} \exp\left[- \frac{(\gamma - 1)^2}{2 \sigma_{\gamma}^2}\right] d\gamma \, dv_0. $$
In Fig.~1 we show various contours corresponding
to different value of $P (\gamma, v_0)$ (excluding the constant pre-factor).  
The probability that $\gamma$ and $v_0$
at the end of a cycle jointly lie within a certain area in Fig.~1 is easily
obtained by integrating  $\int P (\gamma, v_0) \, d\gamma \, dv_0$ over that area. On carrying
out this integration over the shaded region in Fig.~1 that corresponds to
conditions for producing grand minima, we find the value of the integral to be
$0.013$ (i.e., $1.3\%$), remarkably close to the probability of occurrence of grand minima on the basis
of observational data \cite{usos07}. Table~1 gives the value of this integral for a few combinations of 
$\sigma_v$ and $\sigma_{\gamma}$. It may be noted that there would be contributions to the
probability from the regions to the left and to the bottom of the shaded region in Fig.~1.
We have checked that these contributions are negligible.

\vspace{0.3cm}
\noindent
{\bf Table 1.}
{\small This table gives the values of the probability (in \%) of the initiation of
grand minima (given by $\int P (\gamma, v_0) \, d\gamma \, dv_0$ over the shaded region of Fig.~1) for different combinations of the standard deviations $\sigma_v$ 
and $\sigma_{\gamma}$.}

\begin{tabular}{|cc|ccc|}
\hline
 & & & $\sigma_{\gamma}$ & \\
 & & 0.25 & 0.35 & 0.46  \\
\hline
  & 3.20 & 0.8 & 1.3 & 1.7 \\
 $\sigma_v$ &3.34 & 0.9 & 1.3 & 1.9  \\
 & 3.50 & 1.1 & 1.7 & 2.2  \\
\hline
\end{tabular}
\vspace{0.5cm}

\begin{figure*}
\centerline{\includegraphics[width=0.9\textwidth]{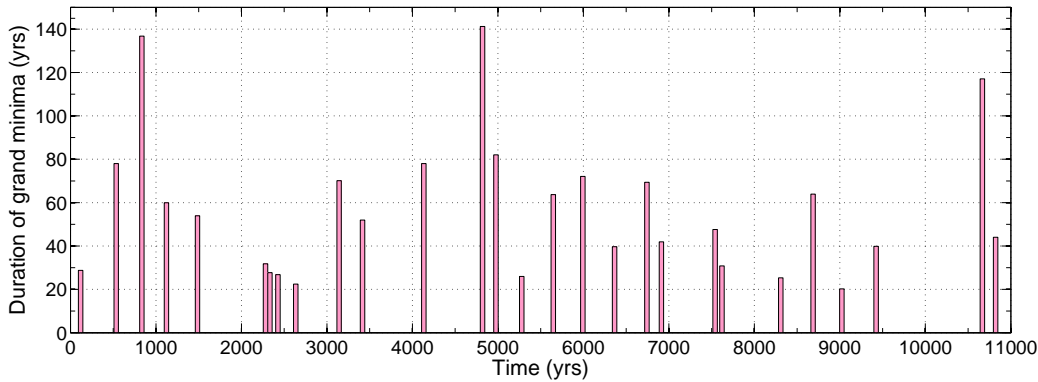}}
\caption{The durations of grand minima indicated by vertical bars at their times of occurrence in a 11,000~yr simulation.
This is the result of a particular realization of random fluctuations that produced 28 grand minima.}
\label{simu_grand}
\end{figure*}

To check whether grand minima really do occur in accordance with the above
simple probability estimate, we have carried out extensive simulations on the
basis of our dynamo code \cite{nc02, cnc04}. 
Karak \cite{k10} changed the values
of a few parameters and our present simulations are based on this model.  We introduce
fluctuations in \pf\ generation by the method proposed by Choudhuri et al. \cite{ccj07}.
At the end of every cycle, we multiply the \pf\ above $0.8 R_{\odot}$
by the random number $\gamma$ obeying the Gaussian distribution shown in 
Fig.~2b. This procedure introduces fluctuations in the poloidal field generated
in the last cycle lying in the upper portions of the convection zone, whereas 
any poloidal field produced at the earlier cycles lying at the bottom of the
convection zone remains unchanged.  While this procedure introduces a 
momentary discontinuity in the field lines at depth $0.8 R_{\odot}$ (see
Fig.~1 in Jiang et al.\ \cite{ccj07}), this discontinuity disappears soon and does not
cause any problem. 
To introduce fluctuations in \mc, we change its amplitude randomly after every
30~yr such that the amplitudes obey the Gaussian distribution shown in
Fig.~2a.  On running the code for 11,000~yr in this way, we obtain the number of grand
minima in the range 24--30 for different realizations of randomly
generated \mc\ and $\gamma$---remarkably close to the observational finding of 27 grand minima in the
last 11,000~yr. Another important result is that we find the Sun to spend 
about 10--15\% of the time in a grand minimum state, which is very close to 17\%
found in the observational study \cite{usos07}.
Fig.~3 shows the durations of grand minima and their times of occurrence for a particular run. 
We ourselves have been amazed that the observational data are reproduced 
so well on running our code with the simple incorporation
of the fluctuations suggested by the histograms in Fig.~2, without having to change
any parameters of the dynamo model compared to our previous work.
Fig.~4 is a sample plot showing how the sunspot number varied for a typical 
1000 years during which two grand minima occurred. Histograms of the
duration of grand minima and waiting time between grand minima are shown
in Fig.~5.  These histograms are to be compared with Figs.~7 and $6$ in Usoskin et al.\ \cite{usos07}.
In the simulations presented above, we have changed meridional circulation abruptly 
after every 30~yr. To check how the above results change with this coherence time $\tau$, 
we have done several simulations with different $\tau$. We see that even when $\tau \sim 15$~yr,
we get around 15--20 grand minima. However, when $\tau$ is less than this value,
the number of grand minima gets very much reduced. Another important point: instead of 
changing \mc\ abruptly at a time, if we change it smoothly in few years, then also
the results do not change significantly.

\begin{figure}[h]
\centerline{\includegraphics[width=0.6\textwidth]{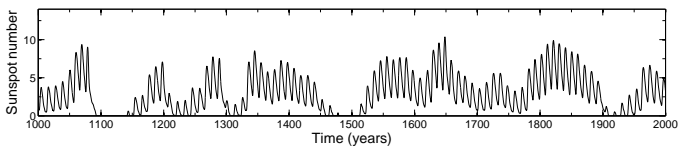}}
\caption{The plot of sunspot number against time for a typical 1000 years. The two
grand minima around 1100 and 1500 years can be seen in Fig.~3 as well.}
\label{simu_grand}
\end{figure}
\begin{figure}[h]
\centerline{\includegraphics[width=0.6\textwidth]{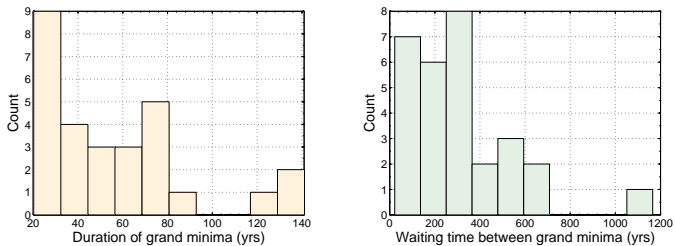}}
\caption{Left panel shows the distribution of the durations of the grand minima 
and the right panel shows the distribution of the waiting times between the grand 
minima.}
\label{simu_grand}
\end{figure}

When sunspots are absent, the Babcock--Leighton process for the generation of
\pf\ cannot take place. Presumably, during a grand minimum, the \pf\ has to be
generated by the $\alpha$-effect originally proposed by Parker \cite{parker55} and 
Steenbeck, Krause \& R\"adler \cite{steen}. It is possible that an $\alpha$-effect coexists along
with the Babcock--Leighton mechanism all the time, although its nature and spatial
distribution (even its sign) is completely unclear at this time.  In view
of this uncertainty, our results presented
in Figs.~3 and 4 are obtained by using the same form of $\alpha$ at all times.
Once the Sun is pushed into a grand minimum, it comes out of the grand minimum
in a time of the order of dynamo growth time, as discussed in detail by Choudhuri \& Karak
\cite{ck09}. One intriguing fact to note is that we get durations of grand minima
similar to their observed values when we keep using the same $\alpha$
throughout the grand minima.
  

We conclude that the irregularities of solar cycles including the grand minima
are produced by fluctuations in poloidal field generation and in meridional circulation.
Assuming these fluctuations to obey Gaussian distributions, we obtain the basic parameters
of the Gaussians from the distribution of the peak sunspot numbers 
and the durations of the last 28 solar cycles.
On running our code with such fluctuations, we find that our theoretical dynamo model
produces grand minima at a frequency remarkably close to what is found in the data for
the last 11,000 yr.

\begin{acknowledgements}
This work is partially supported by JC Bose fellowship of DST awarded to ARC.
BBK thanks CSIR, India for financial support. ARC is grateful to Saku Tsuneta
for the warm hospitality in NAOJ where the revised version of the paper was produced.
\end{acknowledgements}

\end{document}